%% file: main.tex
\documentclass[10pt,journal,compsoc]{IEEEtran}
\usepackage{mathptmx}                  

\usepackage{amsmath,amsfonts}
\usepackage{algorithmic}
\usepackage{algorithm}
\usepackage{array}
\usepackage[caption=false,font=normalsize,labelfont=sf,textfont=sf]{subfig}
\usepackage{textcomp}
\usepackage{stfloats}
\usepackage{url}
\usepackage{verbatim}
\usepackage{graphicx}
\usepackage{cite}
\usepackage{mathtools}
\usepackage{multirow}
\usepackage{color, colortbl}
\newcolumntype{P}[1]{>{\centering\arraybackslash}p{#1}}
\newcolumntype{M}[1]{>{\centering\arraybackslash}m{#1}}

\usepackage{makecell}

\usepackage[svgnames,table,xcdraw]{xcolor}
\definecolor{cb_mint}{rgb}{0.0,1.0,1.0}
\definecolor{cb_blue}{rgb}{0.0,0.0,1.0}

\definecolor{cb_orange}{rgb}{1.0,0.51,0.0}
\definecolor{cb_yellow}{rgb}{1.0,0.8,0.0}
\definecolor{cb_mint}{rgb}{0.0,1.0,1.0}
\definecolor{cb_cerulean}{rgb}{0.0,0.48,0.65}
\definecolor{cb_blue}{rgb}{0.0,0.0,1.0}
\definecolor{cb_lightblue}{rgb}{0.22,0.49,0.72}
\definecolor{cb_green}{rgb}{0.3,0.67,0.29}
\definecolor{cb_red}{rgb}{0.89,0.1,0.11}
\definecolor{cb_purple}{rgb}{0.6, 0.31, 0.64}
\definecolor{cadetgrey}{rgb}{0.57, 0.64, 0.69}
\definecolor{cb_black}{rgb}{0.0,0.0,0.0}

\newcommand{\edit}[1]{\textcolor{cb_black}{#1}}
\newcommand{\new}[1]{\textcolor{cb_black}{{#1}}}
\newcommand{\minor}[1]{\textcolor{cb_black}{{#1}}}

\newcommand{\etal}{\textit{et al.}}
\newcommand{\xrops}{\texttt{XROps}}

\newcommand{\server}{\texttt{Server}}
\newcommand{\device}{\texttt{XR~Device}}
\newcommand{\ui}{\texttt{WebUI}}

\usepackage{ctable}
\usepackage{multicol}

\begin{document}


\title{XROps: A Visual Workflow Management System for Dynamic Immersive Analytics}

\author{Suemin Jeon\thanks{Suemin Jeon is with Korea University. E-mail: orangeblush@korea.ac.kr}, JunYoung Choi\thanks{JunYoung Choi is with VIENCE Inc. E-mail: jychoi@vience.co.kr}, Haejin Jeong\thanks{Haejin Jeong is with Korea University. E-mail: haejinjeong@korea.ac.kr}, Won-Ki Jeong\thanks{Won-Ki Jeong is with Korea University. E-mail: wkjeong@korea.ac.kr}}



\IEEEtitleabstractindextext{
\begin{abstract}
Immersive analytics is gaining attention across multiple domains due to its capability to facilitate intuitive data analysis in expansive environments through user interaction with data. 
%
%
%
However, creating immersive analytics systems for specific tasks is challenging due to the need for programming expertise and significant development effort.
%
%
%
Despite the introduction of various immersive visualization authoring toolkits, domain experts still face hurdles in adopting immersive analytics into their workflow, particularly when faced with dynamically changing tasks and data in real time.
%
%
To lower such technical barriers, we introduce \xrops, a web-based authoring system that allows users to create immersive analytics applications through interactive visual programming, without the need for low-level scripting or coding.
%
%
%
%
\new{\xrops~enables dynamic immersive analytics authoring by allowing users to modify each step of the data visualization process with immediate feedback, enabling them to build visualizations on-the-fly and adapt to changing environments. 
It also supports the integration and visualization of real-time sensor data from XR devices—a key feature of immersive analytics—facilitating the creation of various analysis scenarios.
}
We evaluated the usability of \xrops~through a user study and demonstrate its efficacy and usefulness 
in several example scenarios. 
We have released a web platform (https://vience.io/xrops) to demonstrate various examples to supplement our findings.

\end{abstract}

}

\maketitle

\input{contents/teaser}

\input{contents/Introduction}

\input{contents/RelatedWork}
\input{contents/DesignRequirements}
\input{contents/XROps}

\input{contents/Use_scenario}

\input{contents/Evaluation}
\input{contents/Discussion}

\input{contents/Conclusion}


\section*{Acknowledgments}
This work was partially supported by the National Research Foundation of Korea (RS-2024-00349697, NRF-2021R1A6A1A13044830), the Institute for Information \& Communications Technology Planning \& Evaluation (IITP-2025-RS-2020-II201819), the National Research Council of Science \& Technology grant by MSIT (No. GTL24031-900), the Technology Development Program (RS-2024-00437796) funded by the Ministry of SMEs and Startups, and a Korea University Grant. 



\bibliographystyle{IEEEtran}
\bibliography{reference1.bib}

 
%


 

\vspace{-0.2in}

\begin{IEEEbiography}[{\includegraphics[width=1in,height=1.25in,clip,keepaspectratio]{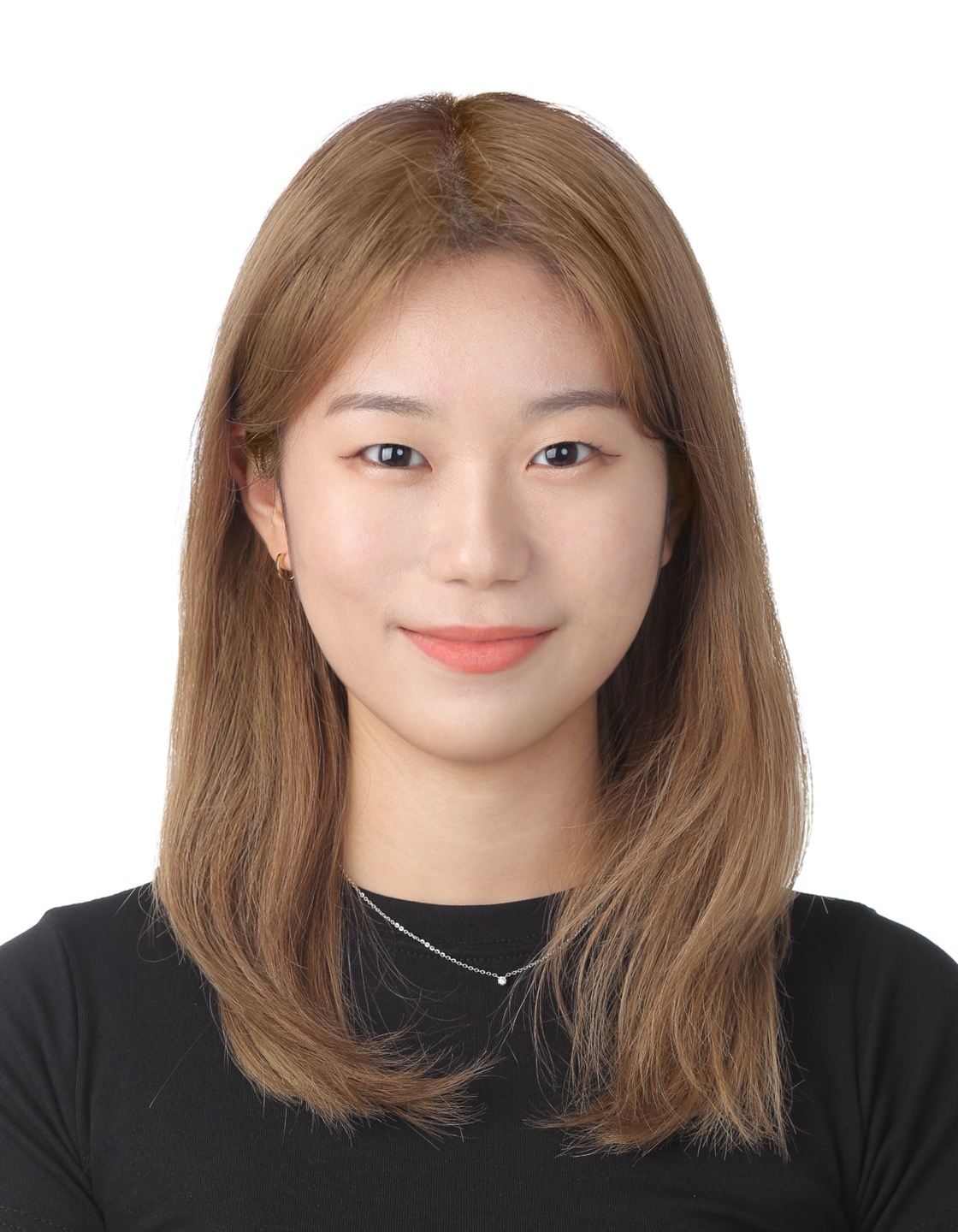}}]{Suemin Jeon} received the BS degree from Korea University in 2022. She is currently working towards a PhD degree with the Department of Computer
Science and Engineering, Korea University. Her research interests include visualization and extended reality.
\end{IEEEbiography}

\vspace{-0.2in}

\begin{IEEEbiography}[{\includegraphics[width=1in,height=1.25in,clip,keepaspectratio]{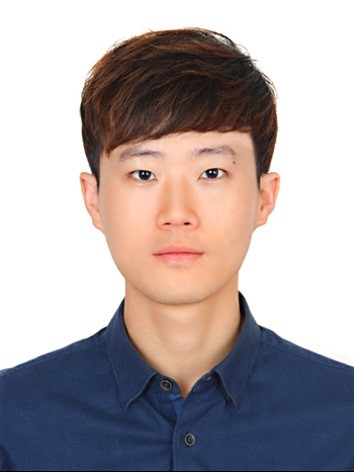}}]{JunYoung Choi} is the Chief Technology Officer (CTO) at VIENCE Inc. He was a Postdoctoral Researcher at Korea University from 2022 to 2024 and received his Ph.D. in Electrical and Computer Engineering from UNIST in 2022. His research interests include visual analytics, visual programming, immersive analytics, and explainable AI.
\end{IEEEbiography}

\vspace{-0.2in}

\begin{IEEEbiography}[{\includegraphics[width=1in,height=1.25in,clip,keepaspectratio]{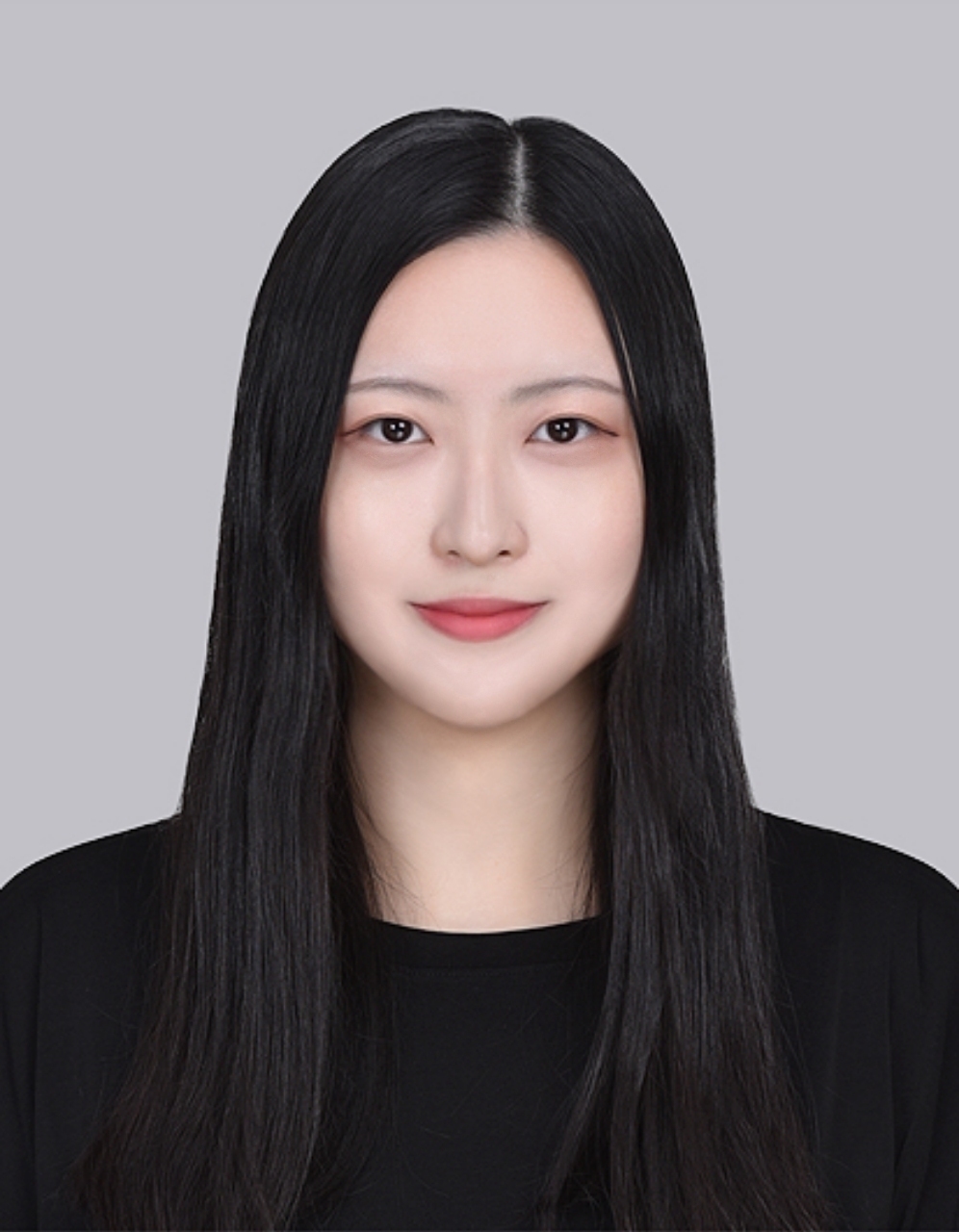}}]{Haejin Jeong} received her BS degree from the Ulsan National Institute of Science and Technology (UNIST). She is currently pursuing a PhD in the Department of Computer Science and Engineering at Korea University. Her research interests include visualization and machine learning.
\end{IEEEbiography}

\vspace{-0.2in}

\begin{IEEEbiography}[{\includegraphics[width=1in,height=1.25in,clip,keepaspectratio]{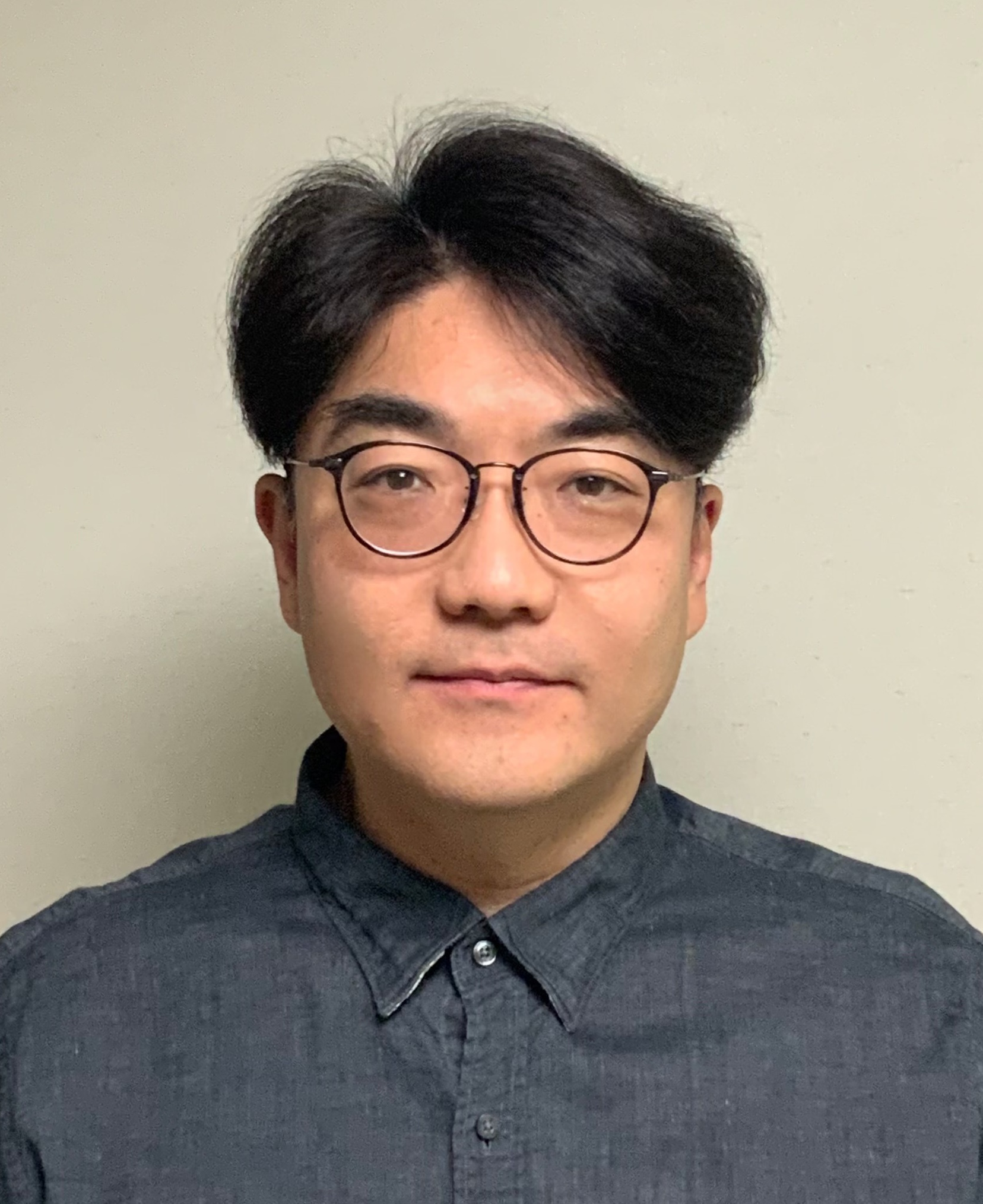}}]{Won-Ki Jeong} is currently a professor in the Department of Computer Science and Engineering at Korea University. He was an assistant and associate professor in the School of Electrical and Computer Engineering at UNIST (2011-2020), a visiting associate professor of neurobiology at Harvard Medical School (2017–2018), and a research scientist at the Center for Brain Science at Harvard University (2008–2011). His research interests include visualization, image processing, and parallel computing. He received a Ph.D. degree in Computer Science from the University of Utah in 2008, and was a member of the Scientific Computing and Imaging (SCI) Institute.
\end{IEEEbiography}





\end{document}

%% file: contents/teaser.tex
\begin{figure*}
  \centering
  \includegraphics[width=1.0\linewidth]{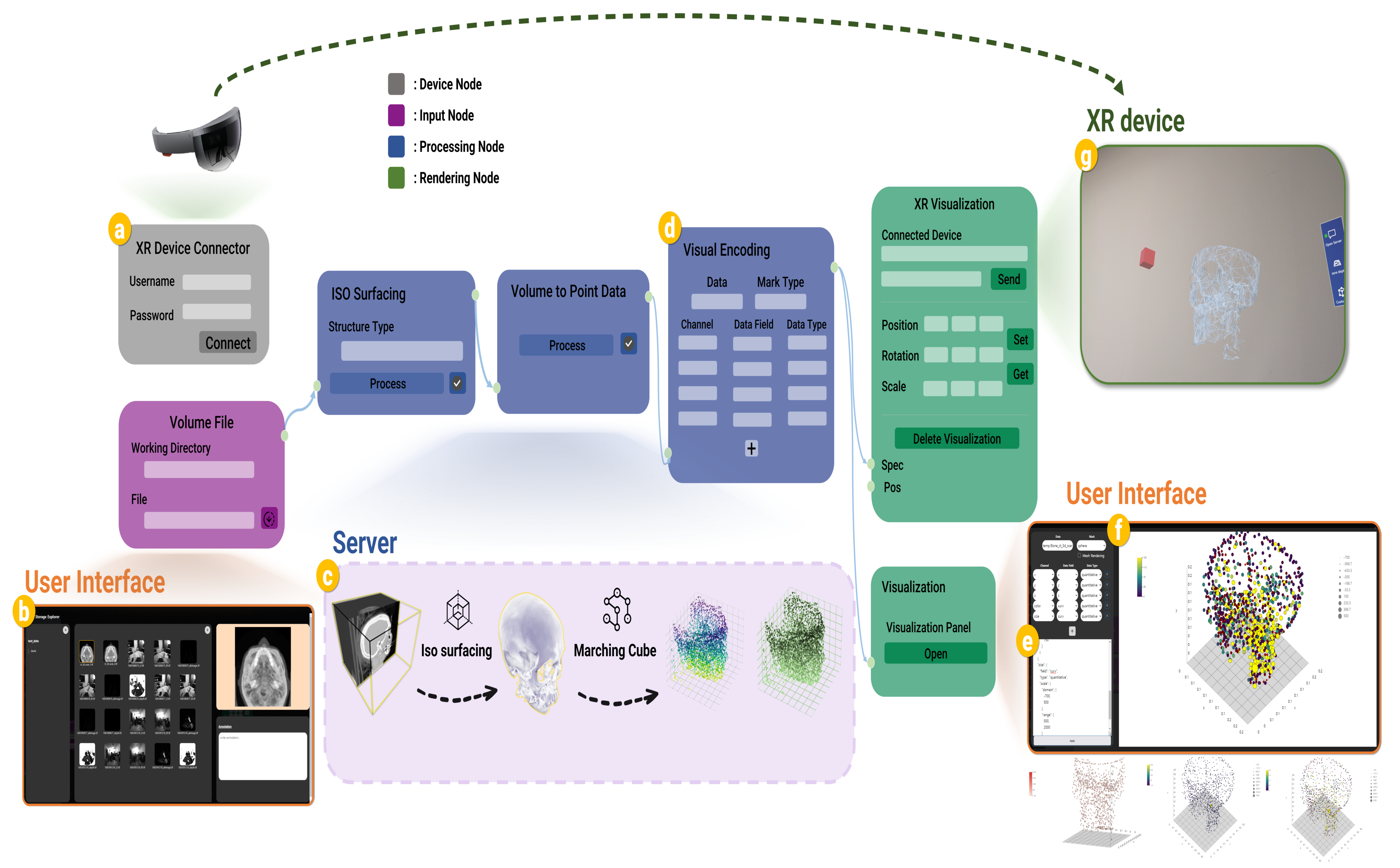}
  \caption{
  An example of \xrops~\new{workflow} 
  %
  for visual exploration of CT data on an extended reality (XR) device (e.g., HoloLens). 
  %
  (a) \xrops~connects the \device~for visualization. (b) CT volume data can be selected using a file browser interface. 
  (c) Several data processing filters, such as ISO-surfacing and Marching Cube algorithms, are applied on the \server. 
  (d) The user can select a preferred visual encoding via simple node interactions or (e) text editing.
  (f) The final visualization can be displayed on the desktop monitor or (g) \device.
  All data processing and visual encoding tasks (b, c, d, and e) can be modified on the fly and instantly displayed (g and f).
  %
  }
  \label{fig:teaser}
\end{figure*}

%% file: contents/Introduction.tex
\section{Introduction}
\begin{figure}
  \centering
  \includegraphics[width=1.0\linewidth]{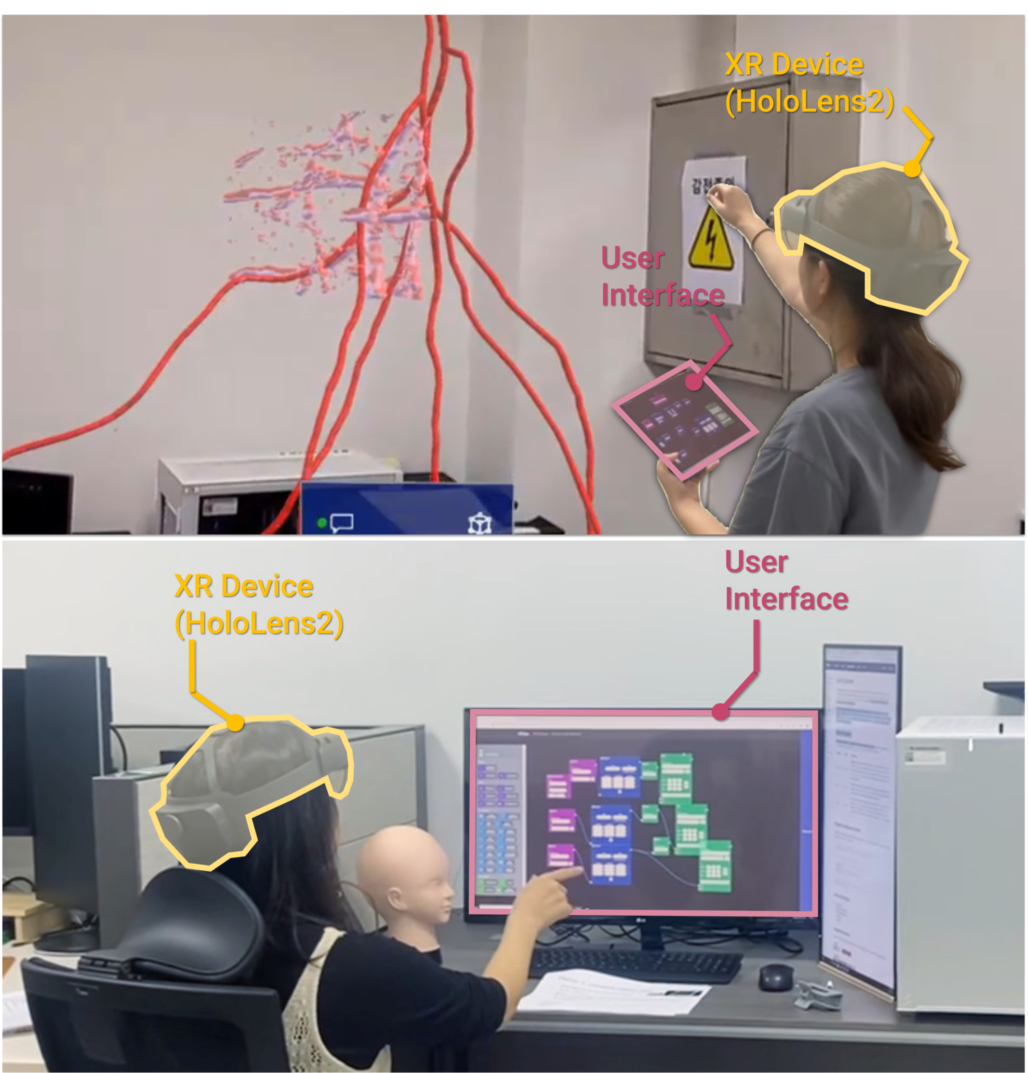}
  \vspace{-15pt}
    \caption{Example of hybrid user interface of \xrops~with tablets and desktop}
  \label{fig:interface}
\end{figure}

\IEEEPARstart{I}{mmersive} analytics (IA) is an emerging research field that leverage user interactions and immersive display technologies, such as extended reality (XR), to support intuitive data analysis.
IA has recently garnered significant attention due to its potential benefits in data exploration and decision-making, offering an expanded 
workspace, intuitive interaction, and realistic 3D perception~\cite{fonnet2019survey}. 
A wide range of studies has applied IA across various domains, including manufacturing~\cite{sahoo2022smart}, sports~\cite{lin2021towards}, biology~\cite{zhang2019biovr}, education~\cite{pellas2020scoping}, healthcare~\cite{qu2022review}, and cultural heritage~\cite{bekele2018survey}.
Despite these advancements, IA remains predominantly a research-focused topic with limited real-world adoption. This limited uptake is largely because existing IA applications tend to be highly domain (or task)-specific and often demand a steep learning curve involving low-level programming skills, hindering broader usage and adaptation to user-specific needs. 
To address these challenges, several IA-authoring toolkits have been developed to reduce technical barriers~\cite{sicat2018dxr,cordeil2019iatk,fleck2022ragrug,rivas2022hiruxr,butcher2020vria}. 
These toolkits enabled visualization designers to create immersive visualizations by writing relatively simple visualization-specific scripts (e.g., Vega-Lite~\cite{satyanarayan2016vega}) or using a graphical user interface (GUI) that do not require extensive low-level programming.
However, most of these toolkits are restricted to generating basic visualizations for specific data types or visualization options, limiting their flexibility for building a complex data processing workflow tailored to specific use cases.
In addition, existing toolkits require significant development to be carried out in advance in an offline environment, which impairs the efficiency of the entire development process. 
In many scenarios, 
the leading XR environment scene development must be performed using Unity~\cite{Unity}, and implementing the necessary functions and interactions requires low-level programming using C\#. 
Subsequently, the code needs to be compiled and  deployed on an XR device for verification.
If modifications are needed, the development cycle must be repeated, making the process time-consuming and inefficient.

\new{Additionally, most current toolkits rely heavily on static setups for data utilization and processing. While these tools are primarily focused on visualization, scene construction, and interaction, they often function merely as consumers of preprocessed data. 
This limits their ability to handle dynamic, real-time inputs. However, \minor{there is a growing demand} for integrating data generated on-the-fly, such as sensor data, in IA scenarios that require real-time data collection and processing. 
Unfortunately, no existing toolkit supports the use of such data at the authoring level, forcing users to manually connect to APIs or make low-code modifications, which poses significant challenges, especially for novice users.}

In response to these challenges, we propose \xrops, a web-based visual programming and workflow management system for authoring IA applications. 
\new{Drawing inspiration from modern software development trends, such as DevOps~\cite{leite2019survey} and MLOps~\cite{alla2021mlops}, which emphasize automation and collaboration to streamline development and operations,
we named our system \xrops.
\xrops~is designed to manage the entire development and execution lifecycle required for immersive visualization, encompassing data collection, processing, analysis, validation, and deployment in dynamically evolving environments, including changes in data sources and tasks. }
To address this, we propose a 
\textit{dynamic workflow management} system that can be easily modified on-the-fly \new{by connecting various \textit{Nodes} for data collection, data processing, visualization, and analysis (Fig.~\ref{fig:teaser})}, providing immediate feedback without the need for time-consuming code revisions and build processes. 
We also leverage 
various sensor inputs collected by XR devices so that the visualization can dynamically adapt to its surroundings. 
In this work, we focus specifically on head-mounted display (HMD) sensors, including depth and RGB cameras, to interact with the environment without using external sensor devices.
The key contributions of this work can be summarized as follows.

\begin{itemize}

    \item We propose a workflow management \new{system} that manages the entire XR visualization development lifecycle. \edit{Since the system can be configured online, dynamic workflow modification is feasible for fast prototyping.} 
    
    \item We introduce a user-friendly web-based visual programming environment for IA development (Fig.~\ref{fig:interface}). \new{The proposed system uses a hybrid user interface, combining a 2D desktop or tablet screen with a 3D XR environment, allowing users to build immersive visualizations easily with familiar interactions.}
    
    \item \new{We designed \xrops~to accommodate a broad range of input data types, including scientific and sensor data, to address a wide array of real-world scenarios. Additionally, we provide abstractions for sensor data processing, as detailed in Section~\ref{sec:sensor}, enabling IA to dynamically adapt to real-time environmental changes.}

\end{itemize}

\new{
We demonstrate the effectiveness and applicability of \xrops~through a user study and various usage examples. 
Our results show that \xrops~enables non-programmers to interactively create immersive visualizations with minimal training. 
\xrops~source code is publicly available (https://github.com/hvcl/xrops)}.

%% file: contents/RelatedWork.tex
\definecolor{HL}{rgb}{0.88, 0.88, 1}
\newcolumntype{h}{>{\columncolor{HL}}c}
\renewcommand{\arraystretch}{1.4}
\begin{table*}[]
\fontsize{7pt}{7pt}\selectfont
\centering
\caption{Comparison of various IA authoring toolkits. \xrops~satisfies the needs of both non-programmers and expert programmers through visual programming and flexible dataflow while accomodating a variety of input data types.}
\label{Tab:comparison}
\resizebox{\linewidth}{!}{
\begin{tabular}{lcccch}
\hline
         \textbf{Feature} & \textbf{DXR} & \textbf{VRIA} & \textbf{RagRug} & \textbf{Wizualization} & \textbf{\xrops} \\          \hline

         Target users & 1 / 2 / 3 & 1 / 2 / 3 & 2 / 3 & 1 / 2 / 3  & 1 / 2 / 3 \\
         \multicolumn{2}{l}{\textit{\fontsize{7pt}{7pt}\selectfont* 1: non-programmer, 2: novice programmer, 3: expert programmer}} & & & & \\          \hline

         Supported data types  & 1 & 1 & 1 / 4 & 1 & 1 / 2 / 3 / 4 \\
         \multicolumn{1}{l}{\textit{\fontsize{7pt}{7pt}\selectfont* 1: tabular, 2: mesh, 3: image/volume, 4: environmental data          }} & & & & & \\          \hline

         Dynamic workflow &  & $\triangle$ & $\triangle$ & $\triangle$   & \checkmark\\
         \multicolumn{4}{l}{\textit{\fontsize{7pt}{7pt}\selectfont* $\triangle$: support reactive behavior}} & &\\ [-1ex]       
         \multicolumn{4}{l}{\textit{\fontsize{7pt}{7pt}\selectfont* \checkmark: real-time interactive modification of workflow }} & & \\          \hline

         Real-time data processing &  &  & \checkmark &  & \checkmark \\ 
         \hline
         Authoring Method & \makecell{\rule{0pt}{7pt} JSON script,\\GUI}  & \makecell{\rule{0pt}{7pt} JSON script}  & \makecell{\rule{0pt}{7pt} Visual Programming \\/JavaScript}  & \makecell{\rule{0pt}{7pt} Voice\\/Gesture}  & \makecell{\rule{0pt}{7pt} Visual \\ Programming}  \\ \hline

\end{tabular}
}
\end{table*}

\section{Related Work}

\subsection{Immersive analytics applications}

\new{IA, a research field that utilize XR technology to create immersive visual analysis environments for data analysis in 3D space, has gained significant interest from both the research community and industry in recent years. 
Qu \etal~\cite{qu2022review} introduced applications that leverage immersive technology in healthcare. 
Kraus \etal~\cite{kraus2022immersive} compared how abstract data can be effectively represented within immersive environments by reviewing various IA applications. Ens \etal~\cite{ens2022immersive} provided insights and interpreting data through spatial and physical interaction in IA, achieved by analyzing multiple applications. 
For more domain-specific IA applications, Ye \etal~\cite{ye2020shuttlespace}~and Lin \etal~\cite{lin2021towards}~proposed IA systems that facilitate more intuitive analysis of ball trajectories in sports using 3D space. Gasques \etal~\cite{gasques2021artemis}~proposed the ARTEMIS, IA system for surgical telementoring, enabling remote specialists to assist during surgery. Boedecker \etal~\cite{boedecker2021using}~proposed an IA system that enables the presentation of preoperative 3D models for effective liver surgery planning. Qu \etal~\cite{qu2020intelligent}~proposed an IA approach for genomic data analysis by combining machine learning techniques with XR technology. 
Those examples demonstrate that IA has the potential to enhance visual analytics by providing users with a more natural, intuitive, and immersive data exploration experience. }

\subsection{Immersive analytics authoring toolkits}
\textbf{Overview:} One of the primary challenges in IA development is lowering the technical barriers for novice programmers and XR designers. Developing IA applications often requires experties in various components, including game engines (e.g., Unity), low-level programming languages, APIs, libraries, 3D coordinate systems, and design principles.
Several IA-authoring toolkits have been developed to address these challenges.
For instance, DXR~\cite{sicat2018dxr} offers a simple declarative language inspired by Vega-Lite grammar to facilitate the generation of visualizations on XR devices. 
DXR also support layered authoring, providing different levels of authoring complexity based on user needs. 
\new{VRIA~\cite{butcher2020vria} generates visual encodings for VR using web-based tools. 
While VRIA benefits from the web's shareable and distributable nature, it is built on WebVR and thus inherits the flexibility and stability limitations of the WebVR platform.}
RagRug~\cite{fleck2022ragrug} enables IoT sensors to provide real-time data for reactive immersive visualizations, primarily for situated analytics.  
\new{RagRug uses visual programming to assist the creation of complex IA workflows; however, it is specialized for situated analytics and directly integrates with Node-RED~\cite{Node-RED}, a platform commonly used for building IoT applications, which requires knowledge in setting up the environments and low-level programming.
Wizualization\cite{batch2023wizualization} is another immersive authoring toolkit that leverages speech and gesture recognition to enable users to interactively create and manipulate visualizations in XR environments. It utilizes a novel "hard magic" system approach for intuitive and rule-based visualization construction. }
Table~\ref{Tab:comparison} compares various authoring toolkits aimed at lowering the technical barriers in IA development. 
While these toolkits reduce programming difficulty to some extent, 
they still require low-level programming or manual interaction to modify tasks or workflows. 

\noindent\textbf{Dynamic workflow modification:}
Interactive visualization authoring, which allows users to modify the visual analysis environment based on feedback from visualizations, is crucial for building a visualizations that meet diverse user requirements and handle dynamically changing setups~\cite{ren2014ivisdesigner, satyanarayan2014lyra, satyanarayan2015reactive}.
Existing XR authoring toolkits only partially support this strategy, with most focusing on visual encoding.
\new{For example, toolkits based on game engines like DXR\cite{sicat2018dxr} and IATK\cite{cordeil2019iatk} 
enable modification of visual encoding via in-situ GUIs after app deployment. 
VRIA\cite{butcher2020vria} offers reactive and straightforward data visualization through VRIA-Builder. Similarly, iARVis\cite{chen2023iarvis} provides a hot-reload feature that detects changes and updates visualizations by reloading modified JSON files containing the visualization configurations after mobile app development. 
However, both VRIA and iARVis are limited by their use of JSON scripting, which constrains the range of modifications and functionality available for creating analytic scenarios. This often necessitates regression to the development phase to make more substantial adjustments. 
For a fully dynamic IA experience, all the components constituting IA should be easily modifiable, including data modification, data re-processing, visual encoding, interaction, and scene configuration.
\xrops~employs visual programming for interactive workflow modification, allowing the entire workflow to be reactive and eliminating the need for separate development and deployment stages. This enables immediate application of changes based on user feedback during ongoing analysis. 
}

\subsection{Mixed immersive analytics}

\new{Despite the advantages of IA, non-immersive analytics (non-IA) remains the dominant form of visual analysis. 
Non-immersive tools benefit from technical maturity and user familiarity, making them well-suited for complex computational tasks~\cite{rivas2022hiruxr,hubenschmid2021stream,jansen2023autovis}.  
These tools are also integrated into existing analysis workflows, reducing the need for domain experts to alter their already established processes~\cite{wang2020towards}. 
Hybrid user-interface-based IA applications aim to combine these benefits with immersive analytics by integrating XR headsets with desktops, tablets, and mobile devices. 
Various methods have been developed to differentiate between in-situ and ex-situ analysis within these hybrid interfaces. 
For example, ReLive~\cite{hubenschmid2022relive} combines an in-situ virtual reality view with an ex-situ desktop view to facilitate the analysis of aggregated data in mixed-reality user studies. 
Similarly, AutoVIS~\cite{jansen2023autovis} uses a desktop for analyzing aggregated data while employing VR for in-situ analysis in automotive user interface (AUI) evaluations. 
Other studies have explored in-situ authoring of immersive visualizations using hybrid interfaces. 
STREAM~\cite{hubenschmid2021stream}, for example, combines spatially-aware tablets in conjunction with AR headsets for multimodal interaction with 3D visualizations. 
The use of tablets and mobile devices expands available display space and enhances mobility. 
Satkowski~\etal~\cite{satkowski2021towards} proposed an in-situ authoring method utilizing mobile devices. 
\xrops~offers a hybrid interface that supports both conventional 2D user interactions on  desktop or mobile displays and immersive 3D interactions on XR devices, leveraging the advantages of both worlds. }

%% file: contents/DesignRequirements.tex
\section{Design Requirements}
\label{sec:requirements}

To determine the requirements for the desired system, we conducted a comprehensive literature review of IA research published in the past five years and examined existing IA authoring toolkits relevant to our work.
%
%
The following is a summary of the requirements derived from this review. 

\vspace{5pt}
\noindent\textbf{R1 - Simple authoring: } 
%
%
%
%
The target users of our system are 
%
domain experts from various application fields 
who may not be familiar with XR programming. 
To encourage such users to integrate immersive visualization into their work, 
the system should offer easy-to-use and intuitive IA authoring methods without requiring 
prior knowledge of specialized IDEs, programming languages, or visualization grammars. 
\new{Several existing authoring toolkits~\cite{sicat2018dxr, butcher2020vria, cordeil2019iatk} aim at novice XR programmers by 
adopting 
visualization specifications, GUIs, and visual programming. }
%
%
%
%
One recent approach~\cite{butcher2020vria, batch2023wizualization} is a web-based platform that emphasizes high accessibility and ease of use. 
%
%
Simple authoring methods allow users to focus on solving domain-specific problems without the burden of software engineering.

\vspace{5pt}
\noindent\textbf{R2 - Adaptable to existing workflow: } 
In many cases, non-immersive analysis has been the primary method for data analysis workflows. 
\new{Previous research has identified several challenges in integrating IA into existing workflows~\cite{jansen2023autovis, wang2020towards, hubenschmid2022relive}. 
For example, Wang~\etal~\cite{wang2020towards} pointed out that existing workflows often require the sharing and exploration of data across multiple users.
Additionally, many users remain more familiar with conventional desktop environments for complex data processing and analysis tasks. 
We also observed that existing IA authoring toolkits are not flexible enough to easily adapt to non-immersive analysis workflows. 
Therefore, the system should support features to lower the barrier to integrate IA 
with existing workflows 
while offering flexible customization options. }

\vspace{5pt}
\noindent\textbf{R3 - Supporting various data types:}
\new{To accommodate a wider range of \minor{use cases}, the system must support various data types for visualization and analysis tasks.
Many existing IA authoring toolkits primarily support tabular data~\cite{sicat2018dxr, cordeil2019iatk, butcher2020vria}. 
Moreover, the growing use of data collected through XR devices in immersive analytical applications highlights the need for systems that can collect and leverage raw sensor data from these devices. 
\minor{For existing IA toolkits to handle unsupported data types, such as sensor data, users are required to rely on additional programming or external libraries, which negatively impacts usability.}}

\vspace{5pt}
\noindent\textbf{R4 - Dynamic workflow management:} 
%
%
%
%
%
\new{The conventional development process for IA applications involves offline development, followed by the build, deployment, and verification stages. 
This process depends on game engine ecosystems, such as Unity or Unreal, for essential features like interaction, sensor data collection, processing, and multi-device connectivity.
Even some IA authoring toolkits require these build-deployment-verification steps~\cite{sicat2018dxr, cordeil2019iatk}, which prevent the incorporation of immediate feedback during analysis. This limits IA's adaptability to dynamic changes in data or tasks, especially in scenarios where users are exploring environments, receiving real-time data from sensors, and modifying data processing and analysis tasks on the fly.  }

%% file: contents/XROps.tex
\section{XROps}
\label{sec:xrops}

\begin{figure}
  \centering
  \includegraphics[width=1.0\linewidth]{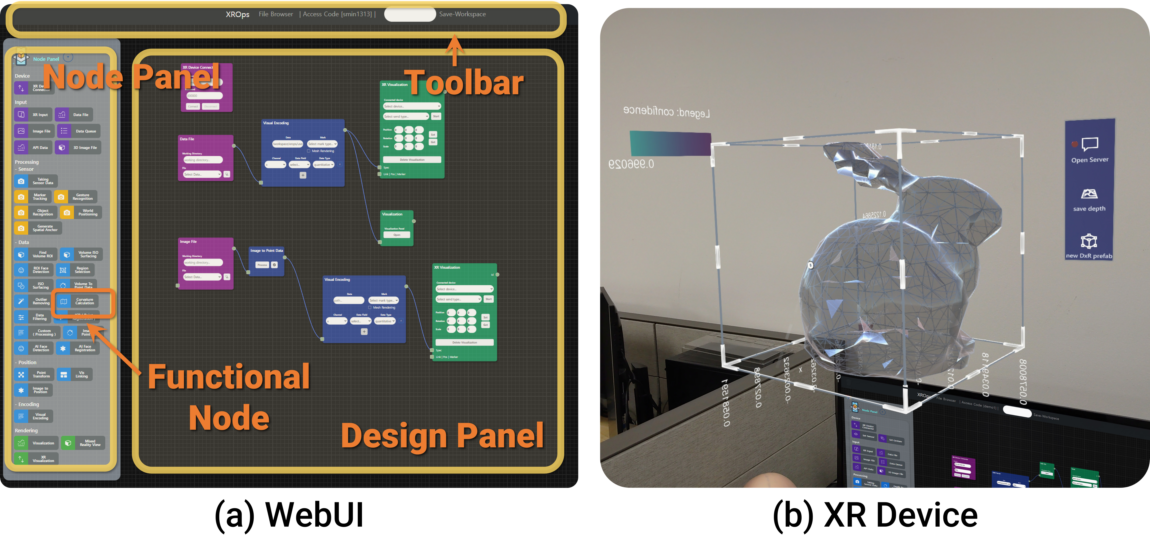}
  \vspace{-15pt}
    \caption{Example views of actual implementation of (a) \ui~and (b) \device~interface.}
  \label{fig:xrops_interface}
\end{figure}

Based on the above design requirements we developed \xrops, a visualization system
%
\new{in which dataflow-based visual programming for workflow management on a client system with a 2D screen (e.g., desktop or tablet) is coupled with interactive visualization on an XR device, creating a hybrid visualization system  (Fig.~\ref{fig:xrops_interface}). 
The proposed visual programming method replaces the traditional XR visualization authoring process by representing scene configuration, sensor data acquisition, data loading and parsing, real-time data processing, visual encoding, and rendering as a dataflow-based workflow. 
This workflow exhibits reactive behavior, enabling on-the-fly workflow modification and immediate visualization feedback on the XR device without 
low-level coding or offline building processes. 
Our current system supports Microsoft HoloLens2 
and HTC Vive devices.} 


\subsection{System Overview}
\begin{figure}[!t]
  \centering
  \includegraphics[width=1.0\linewidth]{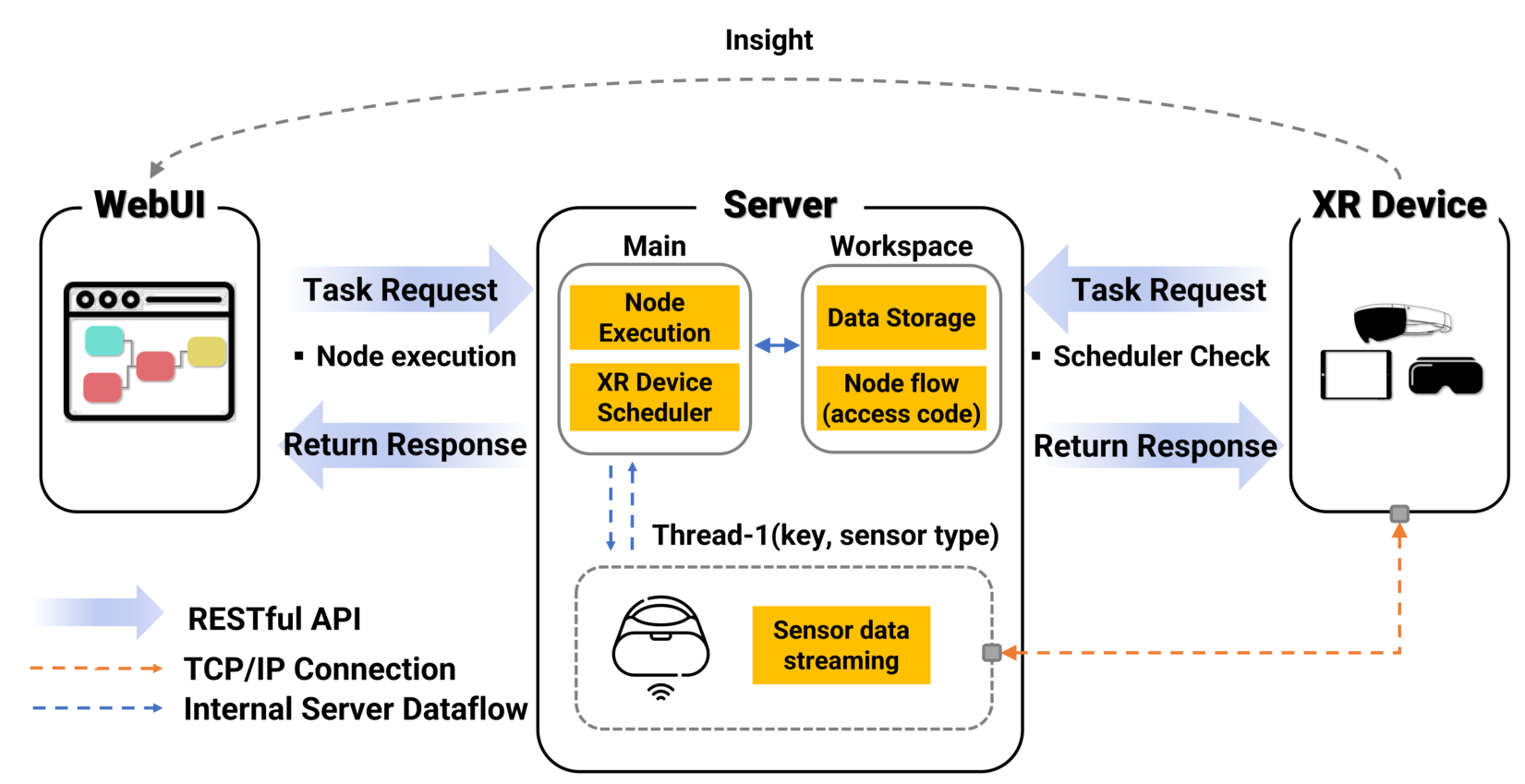}
    \caption{ The \xrops~system overview. The \xrops~system consists of \ui~for creating visualization workflows by connecting functional nodes, \server~that performs the functions of the functional nodes, and \device~responsible for immersive rendering.}
  \label{fig:xrops}
\end{figure}

As shown in Fig.~\ref{fig:xrops}, the \xrops~system comprises three components: \ui, \server, and \device.

%
\vspace{5pt}
\noindent\textbf{WebUI:}
%
\ui~is a web-based interface where users construct an IA system interactively through visual programming.
We chose a web-based 2D user interface for ease of use (\textbf{R1}) and user familiarity (\textbf{R2}). 
%
\new{A web interface eliminates the need to set up a development environment and provides a more familiar user interaction compared to mid-air interactions in an XR environment, which often cause discomfort or 
unintended operations~\cite{hincapie2014consumed}.}
\ui~ includes a node panel, design panel and toolbar as shown in Fig. \ref{fig:xrops_interface}a. 
The node panel contains functional nodes used in visual programming, categorized by their roles. 
Users can place these nodes in the design panel and build workflows interactively by connecting them. 
%
%
The toolbar is used to manage the projects. 
\ui~was developed using JavaScript and React~\cite{boduch2017react} framework, and the Rete.js~\cite{rete} library was used for visual programming. 

\vspace{5pt}
\noindent\textbf{Server:}
The \server~facilitates communication between \ui~and \device~via a network, executing the functional nodes. 
\new{Our client-server design separates user interaction from actual data processing. For example, \ui~and \device~ connect to the \server~via a (wireless) network, allowing users to navigate freely without spatial constraints (\textbf{R4}). 
%
Moreover, it allows more flexible data processing and workflow construction (e.g., resource-demanding and complex data processing is feasible on the server side) and multi-user collaboration (\textbf{R2}).
%
The \server~handles streaming and processing of each device's sensor data through multi-threading. }
%
Communication between the \server~and \ui~is triggered when \ui~requests a specific task mapped to the nodes. 
The \server~then processes tasks such as data handling, device connection, and visualization, and sends results back to \ui~or \device~as needed. 
Dataflow between the \device~and \server~is triggered when \ui~requests sensor data capture or sends visualizations to the \device. 
\new{Sensor data streaming uses a TCP/IP connection for real-time, bidirectional communication with socket connection terminating when streaming ends.}

\new{One common challenge for novice programmers is the initial environment setup. 
To address this, the \server~is built using a Python-based API framework with each function implemented via APIs and containerized using Docker.
This approach simplifies server management and allows users to set up and run the server with just a few simple commands.}

\vspace{5pt}
\noindent\textbf{XR Device:} 
The \device~renders the visualizations created by \ui~and collects the device sensor data. 
%
%
%
\new{It operates based on a pre-built app
, simplifying the development process by eliminating the need to manually set up the XR device or write code to handle sensor data (\textbf{R1, R3}). 
This approach also ensures a more stable performance and minimizes human errors in the code. 
An \device~with the \xrops~app installed can be wirelessly connected to the \xrops~system simply by creating a device node in \ui~(see Section~\ref{sec:device} for more details).}

\begin{figure*}[!bht]
  \centering
  \includegraphics[width=\linewidth]{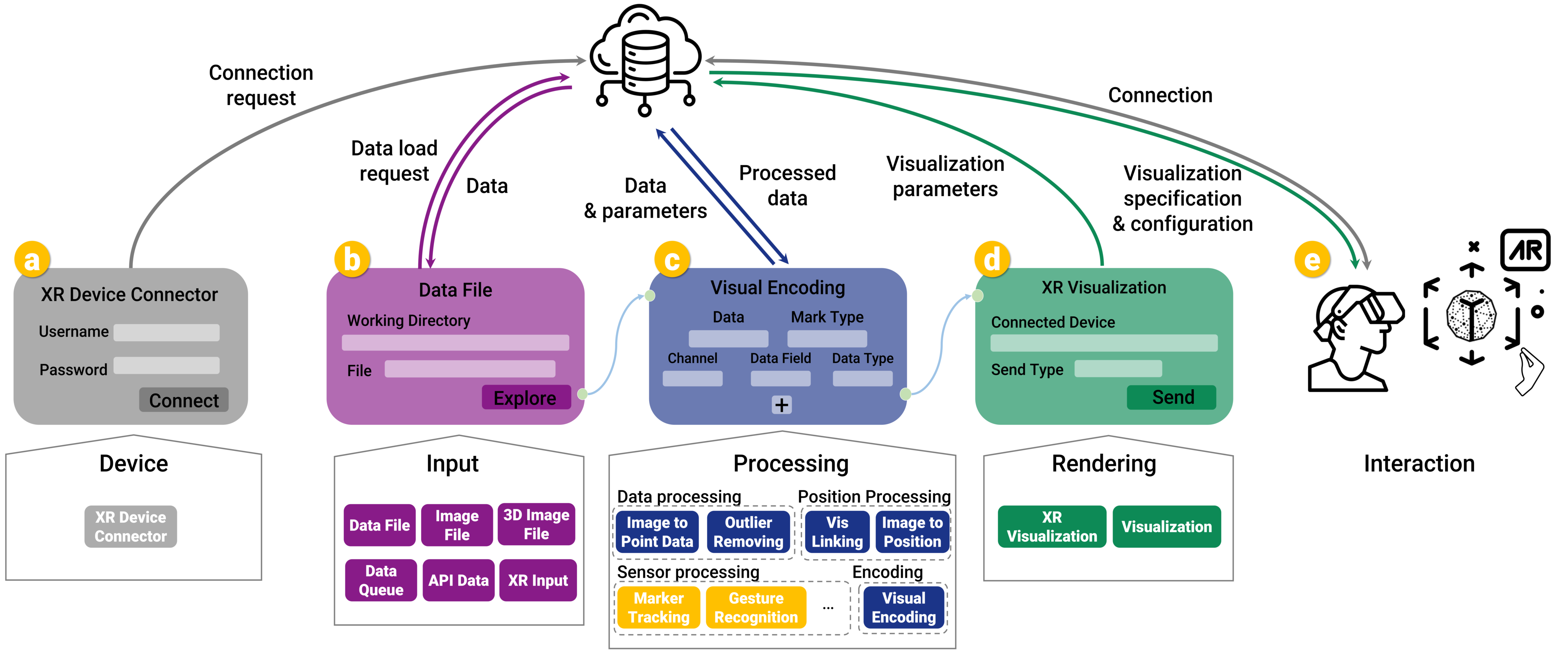}
    \caption{\xrops~workflow. There are four types of functional nodes: (a) Device, (b) Input, (c) Processing, and (d) Rendering, and (e) Interaction on \device. These are connected as dataflow for managing the entire life cycle of XR visualization development.}
  \label{fig:xrops_workflow}
\end{figure*}

\subsection{Workflow Management}

\new{One of the core features of \xrops~is the ability to dynamically modify the data processing workflow through intuitive user interactions (\textbf{R1}, \textbf{R4}). 
\xrops~uses a visual programming paradigm based on a visualization pipeline~\cite{moreland2012survey, wang2016survey}. 
In this paradigm, tasks within the workflow are represented as a directed acyclic graph. 
Users can create and modify this graph to build and adjust the desired workflow.}

A \textit{Node} is defined as an abstraction of a task in the workflow and is grouped into one of four categories based on functionality: \textit{Device}, \textit{Input}, \textit{Processing}, and \textit{Rendering}. 
\new{Processing nodes are further sub-categorized 
into \textit{Data, Position, Sensor}, and \textit{Encoding} types (see Fig.~\ref{fig:xrops_workflow}).}
%
%
%
%
The workflow follows a sequence involving device, input, processing, and rendering nodes, 
and nodes can be connected if their input and output types match. 
For example, users can connect an \device~using a device 
node (Fig. \ref{fig:xrops_workflow}a), load data 
to visualize using an input 
 node (Fig.~\ref{fig:xrops_workflow}b), process the data using a processing
 node (Fig.~\ref{fig:xrops_workflow}c), and display the visualization on the \device~using a rendering 
 node (Fig.~\ref{fig:xrops_workflow}d). 
The visualization can then be viewed and analyzed through interaction on the \device~(Fig.~\ref{fig:xrops_workflow}e). 
Each step is executed through dataflow between the \ui, \server, and \device. 
%
%
%
%
\new{The workflow is re-executed when an event occurs, such as when the user modifies the workflow or when a rendering or sensor node requests a refresh. }
%
%
%
%
\xrops~workflows can be extended, merged, and branched within a workspace, and 
stored on a per-workspace basis. 

\subsection{Node Type Abstraction}

\subsubsection{Device}
\label{sec:device}
%


%

\new{\xrops~provides an \textit{XR Device Connector} node, which manages the connection of the user’s \device~to the \xrops~system.
%
When an \device~sends a connection request to the \xrops~server, the \server~provides the device's assigned username and password.
Once the user inputs this received information into the \textit{XR Device Connector} node,  the \device~is connected to the \server. 
At this point, each \device's information is stored on the \server, and the username acts as a unique device key for handling multi-device functionality. 
This key is used to manage threads for sensor data collection, specifying which XR devices data should be collected from and to which XR devices the data should be sent. }
%
%
%
%
%
Unlike other node types, the device-type node needs to be created and set up only once per device in a single workspace. 

\subsubsection{Input}

Input nodes are responsible for loading or generating data at the application level to be used in later stages of processing or rendering nodes.
The input nodes in \xrops~supports a variety of data types, including tabular, image, volume, mesh, point cloud, and environmental data. 
Once the input data is selected, \ui~requests the data from the \server~and receives it accordingly. 
\new{In the file explorer, only data corresponding to the selected node type can be viewed and loaded, allowing users to preview and select file names from the directory. Additionally, among the input node types, an \textit{XR Input} node generates and passes the device key to subsequent nodes. While the device key is not part of the default \xrops~data type, it is used internally within the sensor data processing nodes.}
To handle accumulated environmental data effectively, users can utilize  a \textit{Data Queue} node. 
The data queue node can store continuously generated sensor data in a queue for later use (saved with the current workspace), or it can be directly connected to the other workflows as input. 
%
%
%
%
\subsubsection{Processing}
\label{sec:processing}

\new{Processing nodes are further classified into four groups: data processing, position processing, sensor  processing and encoding groups. }
%

\vspace{5pt}
\noindent\textbf{Data processing:} 
\new{The data processing node handles tasks such as data type conversion, extraction, and merging across the various data types in \xrops. 
This functionality enables the use of dynamic, live-computed, or interactively created data within the analysis process~\cite{saffo2023unraveling}. 
It facilitates analyzing data in new ways, as data previously available only in static or pre-computed formats can now be processed in real-time using \xrops. 
Data processing nodes can be arranged sequentially to create more complex workflows. 
The actual processing tasks are executed on the \server~according to the specific processing request. 
Additionally, users can create custom data processing nodes beyond the default ones, as detailed further in Section~\ref{sec:custom}.}

\begin{figure}
  \centering
  \includegraphics[width=1.0\linewidth]{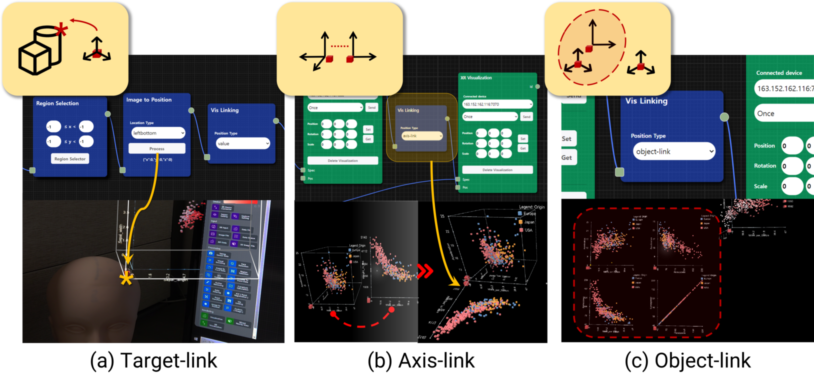}
  \vspace{-20pt}
    \caption{
    Immersive visualizations can be positioned via three different link types. 
    (a) With the target-link type, visualizations can be anchored to the target position. (b) With the axis-link type, the axes of visualizations can be linked. (c) With object-link type, visualizations can be grouped. }
  \label{fig:link}
\end{figure}

%
\vspace{5pt}
\noindent\textbf{Position processing:} 
\new{
The position processing node generates position data from the input data source, which is then used in the rendering node through three types of links, as illustrated in Fig.~\ref{fig:link}.
For the target-link (Fig.~\ref{fig:link}a), the processing node computes the real-world position primarily based on processed sensor data. 
This allows users to configure \minor{visual analytics that leverage spatial information,} using specific points derived from the position processing node. 
The Axis-link (Fig.~\ref{fig:link}b) and Object-link (Fig.~\ref{fig:link}c) are established through the \textit{Vis Linking} node, allowing for the expression of relative relationships with existing visualizations. 
The Axis-link facilitates axis-based analysis by aligning axes with the same data field across two visualizations, while the Object-link enables users to group visualizations.
}

\begin{table*}[]
\fontsize{8pt}{8pt}\selectfont
\centering
\caption{Sensor processing node abstraction based on usage scenarios.
    The table presents the application examples, usage explanation, and input and output types for each sensor processing node type.}
    \label{fig:sensor}
  \includegraphics[width=1.0\linewidth]{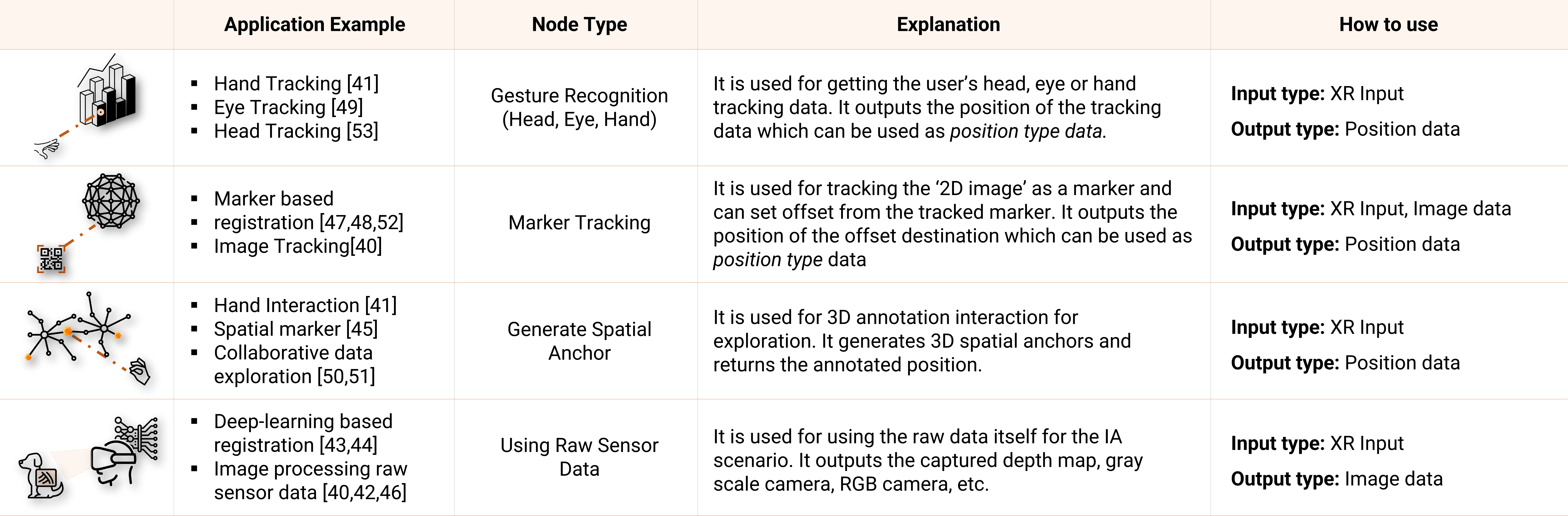}

\end{table*}
\nocite{pei2019wa, romat2020dear,refaey2019eye,hoster2023style,ran2016infocarve,whitlock2019designing,jiang2020augmented,whitlock2020hydrogenar,gutierrez2018phara,mahfoud2016gaze,borhani2022dc,feng2023comprehensive,adagolodjo2018marker,mathews2021air}


%
\vspace{5pt}
\noindent\textbf{Sensor processing:} 
\label{sec:sensor}
%
\new{
The sensor processing node collects sensor data from an \device~and processes it \minor{for use in other nodes}.
%
Sensor data obtained from devices are comes in either image (color and depth camera data) or tabular (head, eye, and hand tracking data) formats.
To meet the design requirements (\textbf{R1}, \textbf{R3}), node abstractions for sensor data processing are provided based on various scenarios derived from a literature survey on IA applications, which are classified into four types (Table~\ref{fig:sensor}).}
%

%
\new{The first type enables IA scenarios creation using user-tracking data from the XR device.
The \textit{Gesture Recognition} node uses head, eye, and hand tracking data as position data, triggering visualizations based on the user's movement or tracking info, directly usable for the rendering node.
The second type, 
%
%
the~\textit{Marker Tracking} node sets desired markers using image files, acquiring their spatial location in real-time for visualizations.
This is useful for marker-based registration or tracking moving targets.
The third type, the \textit{Generate Spatial Anchor} node, creates  spatial anchors from the world coordinates of the IA application.
These anchors store absolute positions and pass them to the next node, enabling 3D annotations and shared views in collaborative analysis. 
Lastly, the \textit{Using Raw Sensor Data} node provides access to raw sensor data, such as depth maps, RGB cameras, and tracking data from XR devices like Microsoft HoloLens2. 
It streams data in real-time or captures moments via air tap or button interaction, selecting desired sensor streams for custom processing nodes.
%
}

%
\vspace{5pt}
\noindent\textbf{Encoding:} 
The encoding group receives processed data from the data processing group and constructs visual encoding. 
Since we uses a modified DXR for \device~visualization, the DXR GUI is implemented in node format, and more specific visual encoding can be customized through the text editor. 
The designed visualization spec is sent to the \server~and passed to the rendering node. 
\new{All data must pass through the visual encoding node before being used as input for rendering node.}
For \device~rendering, users specify the visualization's position via position input, which can be a specific location in the world coordinate system or a relative position linked to pre-rendered visualization. For relative positions, the \textit{Vis Linking node} connects the rendering type of any pre-rendered visualization.
%
%



\subsubsection{Rendering}
Final visualization can be displayed on either a 2D desktop/tablet screen or a 3D \device~via the rendering node. 
%
%
\new{Web-based rendering uses WebGL~\cite{parisi2012webgl}~and Vega-Lite~\cite{satyanarayan2016vega}, allowing visualizations to appear on a 2D screen through a web browser based on visualization \minor{specification}.}
For \device~rendering, the \server~sends the visualization spec to the \device, parsed through the modified DXR.
Position inputs for XR rendering include target value (Fig.~\ref{fig:link}a), axis-link (Fig.~\ref{fig:link}b), and object-link (Fig.~\ref{fig:link}c).
%
%
Multiple devices can receive and display visualizations by adding additional rendering nodes to the workflow, enabling visualization sharing across device.

\subsubsection{Interaction}
\begin{figure}
 \centering
 \includegraphics[width=1.0\linewidth]{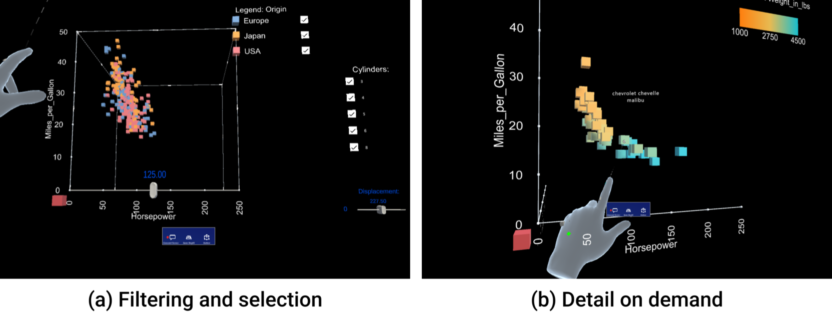}
 \vspace{-20pt}
 \caption{\edit{Interaction examples for data manipulation by \device. \xrops~provide filtering and selection on axis, legend, or field through toggle and threshold filters. \xrops~allows for detailed on-demand information retrieval for fields.}}
   \label{fig:interaction}
\end{figure}

\new{Interaction is task abstraction within the \xrops~workflow (Fig.~\ref{fig:xrops_workflow}e), not a standalone node type.
%
%
Simple user interactions, like hand gestures, can modify visualization parameters after initial rendering, enhancing data understanding.}
%
The interaction box allows users to rotate, scale, and move the visualizations using hand gestures.
Transformation parameters can also be adjusted via the rendering node in \ui.
\xrops~supports various 3D interactions for manipulating data visualization, including filtering, selection, and detail-on-demand, as shown in Fig.~\ref{fig:interaction}. 
%
\new{Hand, head, and eye movements can be integrated into workflows by tracking data from sensors for more immersive interaction.}

\subsection{Execution Management}


\begin{figure}[!tb]
 \centering
 \includegraphics[width=1.0\linewidth]{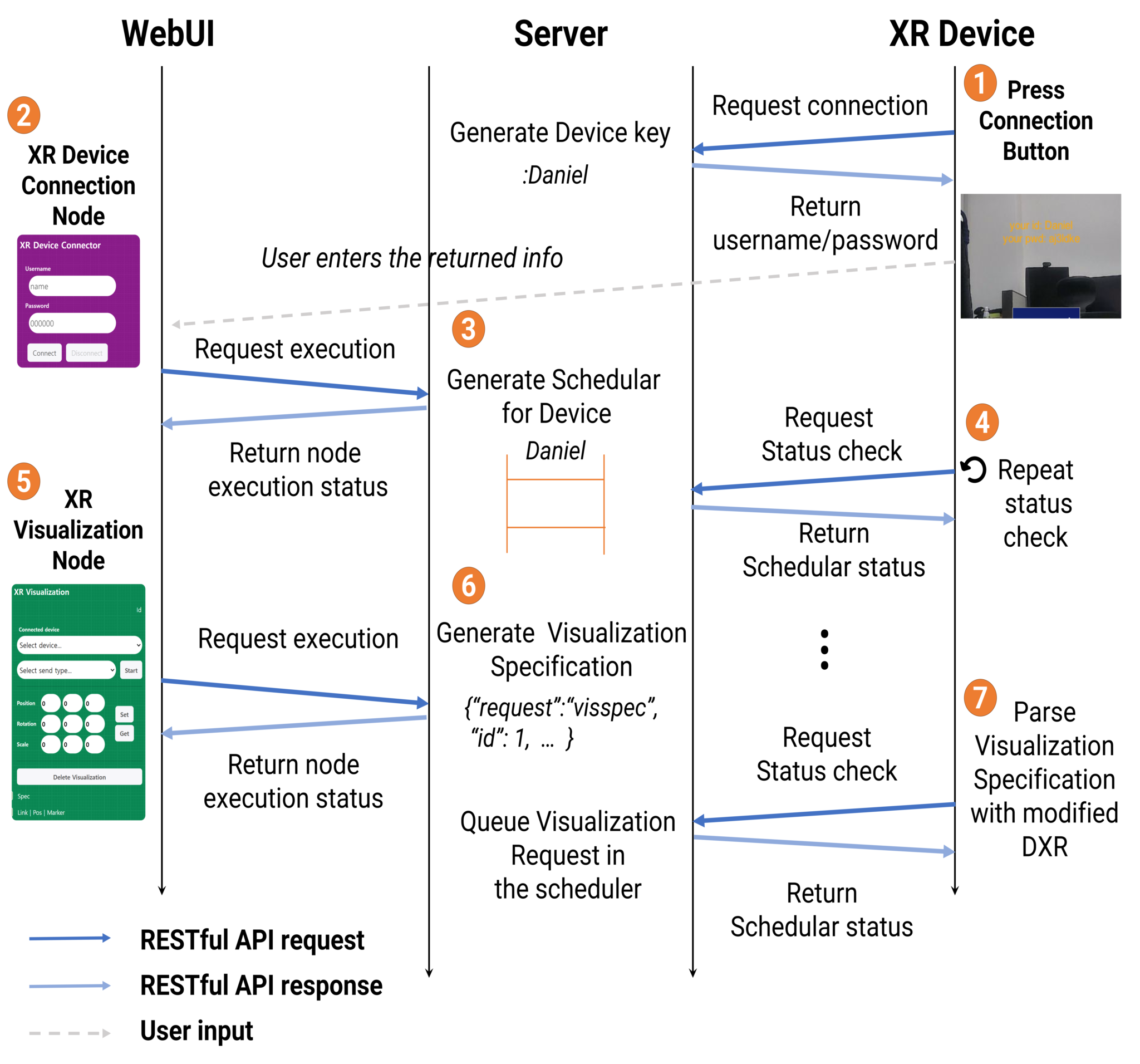}
 \vspace{-10pt}
   \caption{\new{This figure illustrates an example of node execution flow between the \ui, \server, and \device. It shows the exchange of information between three components over the network when the \device~is connected to \xrops, and the visualization is rendered on the \device~ with \textit{XR Visualization Node} execution.}}
 \label{fig:exe}
\end{figure}

\new{\minor{To execute a node}, \ui, \server, and \device~communicate 
via a RESTful API over HTTP. 
Execution can be triggered through user interaction, by changing the node's state, or automatically when a preceding node's state changes. 
Once executed, the node sends a RESTful API request to the \server, which processes it via a corresponding Python function. The function's output is sent back to the web interface. Each \device's scheduler, created when the device connects to the \server, regularly checks for assigned tasks by periodically sending requests to the \server. }
%
%
%
%
%
%
\new{Figure~\ref{fig:exe} shows an example of a node execution flow between \ui, \server, and \device, illustrating how information is exchanged when the \device~connects to \xrops~ and visualizations are rendered. 
\minor{(See Supplementary Fig.~S1 for more detailed explanation.)} 
When a new \device~connects to \xrops, a node sends a request to the \device~for visualization.
First, pressing the connection button on the \device~sends a URL request via the RESTful API (Fig.~\ref{fig:exe}, \textcircled{1}).
The \server~generates a device key and responds with credentials. The user inputs this information into the \textit{XR Device Connection} node via \ui, then \minor{presses} "connect" (Fig.~\ref{fig:exe}, \textcircled{2}). This triggers a node execution request sent to the \server~with the credentials as URL parameters. The \server~verifies the connection, creates a scheduler, and returns the execution status to \ui~(Fig.~\ref{fig:exe}, \textcircled{3}). The \ui~return value usually indicates success or provides the \server's data path where processed data is stored. These return values can be used to modify the node's state or passed as parameters to subsequent nodes.
The \device~periodically sends status check requests to the \server~(Fig.~\ref{fig:exe}, \textcircled{4}), which returns the scheduler’s status. If no task is assigned, the \server~responds with "empty" in JSON format. When a node sends a visualization request (Fig.~\ref{fig:exe}, \textcircled{5}), it includes the device key and visualization details (e.g., transformation, data type). The \server~generates a visualization specification in JSON format and queues it in the scheduler (Fig.~\ref{fig:exe}, \textcircled{6}). The \device~retrieves the specification during its status check (Fig.~\ref{fig:exe}, \textcircled{7}) and executes or visualizes the data accordingly.
%
}

\subsection{Layered Authoring and Customization}
\label{sec:custom}


\xrops~offers layered authoring \minor{(see Supplementary Fig.~S2)} 
to accommodate users with different levels of programming expertise. 
For non-programmers, the framework provides pre-designed workspaces that can be easily customized through simple UI interactions. 
Users with more knowledge of visualization and visual programming can create custom visualization workflows by connecting pre-defined functional nodes, allowing them to tailor visualizations to their specific needs.
Programmers with low-level coding experience can develop custom nodes to further extend the framework's capabilities. 

\new{The Custom node allows users to run their own code as independent nodes within \xrops, addressing specific challenges (\textbf{R2}). As a processing type, Custom nodes \minor{let users connect} to their desktop or cloud servers via an IP input. The \server~retrieves a list of user-provided Python functions, which can be selected within the node. Users can specify and modify parameters for these functions through the node's input field. The selected function is also viewable in text format via \ui.
%
%
This approach facilitates the use of data and code from users’ local environments while preserving the confidentiality of sensitive information.}
%
%

\subsection{Collaboration}
%
%
%
%

\new{
Collaboration in \xrops~occurs in two ways: 
within a single workspace via multi-device and through workspace sharing.}
\new{For multi-device connections, several devices can connect simultaneously to a single workspace using multiple \textit{XR Device Connection} nodes. 
As discussed in Section~\ref{sec:device}, each device is assigned a unique key, which the rendering node uses to route visualizations to the correct device.
%
This allows users to employ multiple rendering nodes, enabling the same IA view to be shared across multiple XR devices. 
Users can also share dynamic environmental data using \textit{Sensor type} nodes, facilitating the exchange of not only static views but also live scenes.
}
\new{
Workspace level collaboration involves editing the same workspace from different locations while rendering the same scene.
To meet design requirement \textbf{R4}, traditional build, deploy, and distribution process are replaced with rendering nodes and workspace management. 
Workflow data is stored in the \server's database on a per-workspace basis, distinguished by an access code. 
This access code acts as a project unit, updating whenever the workspace is saved, allowing shared programming work through code sharing.}
%
%

\subsection{XR Rendering}

\xrops~uses a modified version of the DXR toolkit for XR rendering. 
DXR was modified at the script level to support network rendering and new data types in \xrops. 
\new{In DXR, each data point is treated as a \textit{mark} class containing channel, size, position, color, and direction. 
Every data point goes through an `infer' stage before being `constructed' and `placed'. 
For \xrops, efficient handling of new data types like mesh, volume, and image was essential due to their size. 
Additionally, since real-world data is used, data values often correspond to world coordinate positions, requiring real-time visualization. 
The constantly changing data values and the encoding of content through visual programming required continual updates of visualizations on the XR device via networking.}
\new{To address these needs, a new mark class was created to handle mesh, volume, and image as a single mark, and a \textit{coordinate type} was added to the existing data types to align visualizations with world coordinates.}
\new{
%
For networking purpose, the data for visualization was parsed within the visualization specification and transmitted to the \device, enabling real-time updates using DXR. 
}




%% file: contents/Use_scenario.tex
\section{Use-case Scenarios}
\label{sec:usecase}

\begin{figure*}[!hbt]
  \centering
  \includegraphics[width=1.0\linewidth]{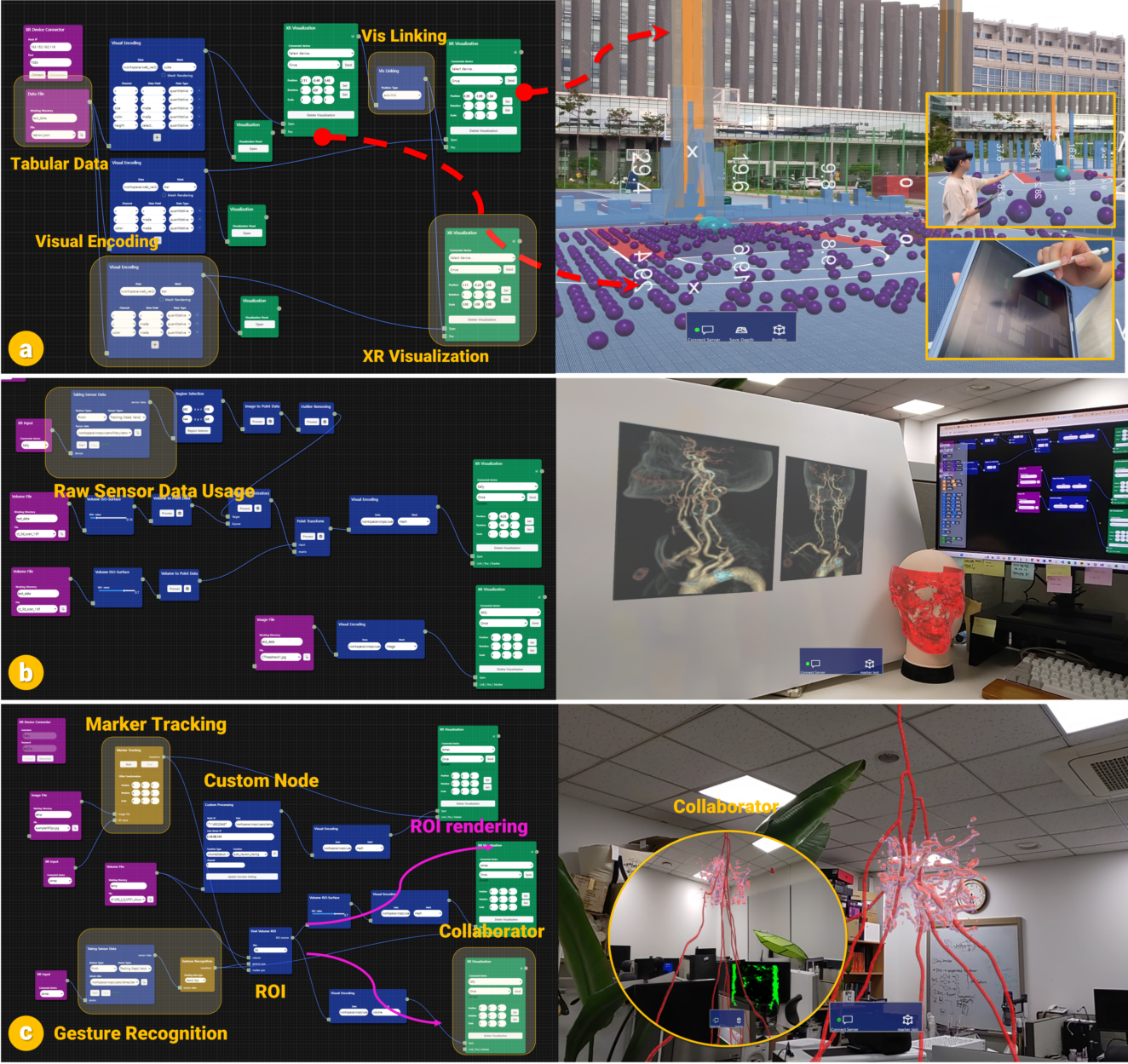}
  \vspace{-15pt}
    \caption{\minor{Use-case scenarios of \xrops~in (a) sports data analysis (b) surgical guidance, and (c) biological data analysis. (a) Sports data, such as a basketball player's shot location and frequency, can be analyzed by linking multiple immersive visualizations. The visualization linking function enables axis-aligned analysis.(b) The CT volume data can be overlaid on the patient's face by utilizing environmental data. (c) The 3D microscopy image of a mouse brain tissue is being collaboratively explored and analyzed. }}
  \label{fig:scenario}
\end{figure*}

%

\minor{This section demonstrates how \xrops~can be applied across various scenarios, ranging from basic examples in immersive analytics to advanced real-world applications. Each case includes a \textit{default} scenario followed by a \textit{dynamic modification} scenario. \minor{A total of four scenarios were explored, with three presented in this section and one is included 
in the supplementary material.}}
%
Interactive demos can be found on the \xrops~website at https://vience.io/xrops (access code: demo1, demo2, demo3, demo4).

\subsection{Sports data analysis (S1)}


\noindent\textbf{Scenario}: Sports data analytics is becoming more prevalent in its application for a range of purposes, such as training, strategy analysis, and evaluating player performance. Here we analyze LeBron James' shot attempts and successes on basketball court~\cite{lebron}~as shown in \minor{Fig.~\ref{fig:scenario}a}. 
%

\noindent\textit{Implementation}: First, a \textit{Data File} node is created in \ui~and the \textit{lebron.json} data file is uploaded. 
A \textit{Visual Encoding} node maps shot attempts and successes to the court's x and y locations, using color and size respectively. 
%
This is connected to an \textit{XR Visualization} node and sent to the \device~to create a scatter plot that enables analysis of LeBron's shot statistics 
on the court plane. 
To further analyze the shots along the court's x and y-axes a bar chart \textit{Visual Encoding} node can be added to the workflow.

\vspace{5pt}
\noindent\textbf{Dynamic modification}: \new{Upon the above analysis feedback, we were able to identify where players have a high success rate and a high number of attempts with precise numerical coordinates. Now we aim to map the player's shot attempts and successes onto the actual basketball court to gain real-world insights.}

\noindent\textit{Implementation}: \new{The `axis-link' feature in the \textit{Vis Linking} node maps axes with shared data fields. Consequently, a scatter plot aligned with the court's x and y coordinates can be placed on the floor, while bar charts linked to these axes can be mapped accordingly using the  \textit{Vis Linking} node.
Visualization can be scaled, and image markers can be positioned at the court's corners and tracked using the \textit{Marker Tracking} node.
Alternatively, raw depth map sensor can be utilized to calculate the real-world position of the court corner. This information can then be transferred as position data for the visualization created above.
%
\xrops~also enables users to modify workflows with a web connection, allowing them to adjust pre-built workflows, like the one depicted in Fig.~\ref{fig:scenario}a, using a tablet or mobile phone, whether indoors or in mobile environments.}

\subsection{\textbf{Medical data analysis (S2)}}
\label{sec:scenairo3}

\noindent\textbf{Scenario}: IA is gaining significant attention in medical data analysis for surgical preparation, training, and guidance due to its ability to provide an intuitive visualization of complex data and enable direct interaction. 
This example showcases a surgical guidance scenario for facial bone surgery using \xrops~and utilizing environmental data. 
XR-based surgical simulations and guidance are actively researched, and facial bone surgery is an example of surgical guidance that is assisted by CT data during surgery. 
The task involves real-time registration of pre-acquired CT scans with patient's face.
For this, we reproduced the work by Gsaxner~\etal~\cite{gsaxner2019markerless} on the face-neck surgery AR surgical guidance scenario using \xrops, as shown in \minor{Fig.~\ref{fig:scenario}b}. 
As a proof-of-concept implementation, this example is not fully automatic but includes essential functions like point cloud generation from CT data, depth map region selection, point cloud registration with iterative closest point (ICP) algorithm, and CT volume rendering onto a real-world background.

\noindent\textit{Implementation}:
%
%
%
First, the CT data is uploaded, and bone and skin point clouds are extracted. This is achieved by loading raw CT volume data via the \textit{3D Image File} node and connecting the \textit{Volume ISO Surfacing} and \textit{Volume to Point Data} nodes. 
Following this, a depth map of the patient's face is captured by the sensor camera, and the face region depth map is converted into point cloud data. 
%
%
%
Currently, the face region is manually selected through the \textit{Region Selection} node, though automation using a custom node with an automatic face region detection library is possible.
The skin point cloud from the CT volume and the patient's face point cloud are then registered using the ICP algorithm.
A transformation matrix is extracted to align the bone mesh with the CT data, leading to the final rendering of the bone mesh overlaid on the patient's face.
%
%
This workflow can be pre-designed and loaded for different surgeries, allowing real-time rendering without interaction during the surgery. 

\noindent\textbf{Dynamic modification}:
\new{After verifying CT data registration with the patient, further analysis of the surgical region using angiography on coronal and sagittal planes may be required. 
%
Additionally, depth visualization from the skin to each point in the CT bone data might be necessary.
To meet these needs, the following tasks can be completed using \xrops.}
%

\noindent\textit{Implementation}: \new{To visualize angiography results, upload data via the \textit{Image Data} node, specify the mark as "image" in the \textit{Visual Encoding} node, and render it through the \textit{XR Visualization} node. Some additional surface features can be visualized as a color map on the mesh (for example, a \textit{Curvature Calculation} node can be used for computing curvature and assigning color values on the mesh). 
For detail-on-demand interaction, the "text" channel can be added with other mark types selected in \textit{Visual Encoding} node. This interaction aids in understanding patient specific details in targeted areas.}

\subsection{Biological data analysis (S3)}
\label{sec:scenairo4}

\noindent\textbf{Scenario}: 
\new{
IA has the advantage of being able to view data in three dimensions and facilitates collaborative data analysis, making it useful in analyzing large-scale scientific data~\cite{liu2021narrative, wang2019vision}. 
This example demonstrates how \xrops~performs explanatory analysis of mouse brain connectivity using large-scale 3D microscopy data, 
as shown in \minor{Fig.~\ref{fig:scenario}c}. 
Due to the difficulty in uploading and visualizing large datasets in IA, we focus on identifying Regions of Interest (ROI) for volume visualization, followed by \minor{data proofreading.}}
%
%

\noindent\textit{Implementation}: \new{First, the confocal microscopy volume data, with a size of 4GB, is uploaded to \xrops~using the \textit{3D Image File} node.
Then, a \textit{Custom} node is connected to the user's cloud server and set the function type to `volume to tabular'. 
Next, the \textit{auto neuron tracing function} from the custom function lists is selected, 
%
which uses a neuron tracing algorithm that returns a list of line segments (e.g., center locations and width of the neuron) with traced neuron structures as 3D mesh for rendering on the XR device. 
To establish a reference between the real world and the IA scene, a marker is placed and tracked using the \textit{marker tracking node} in \xrops. 
%
%
For a detailed examination of ROI, one can use the \textit{Gesture Recognition} node to capture locations via an air tap action. 
The ROI volume is computed by calculating the current mesh's world transformation data relative to the air tap position.
%
%
The ROI volume and traced neuron data are visualized through the visual encoding node, allowing inspection of untraced structures in the area of interest.}

\vspace{5pt}
\noindent\textbf{Dynamic modification}:
\new{After tracing neurons and selecting ROI, the next step is to enable collaborative analysis by sharing the ROI with multiple participants. }

\noindent\textit{Implementation}: 
\new{For collaborative analysis, the reference marker is placed, and the device key 
is changed to that of the new user to share the scene with the collaborator. 
If the new collaborator wants to analyze the ROI using different visualization, they can add a new branch in the workflow after the \textit{Find Volume ROI} node. 
Furthermore, if one wishes to perform collaborative analysis via 3D annotation, create a new workflow using a \textit{Generate Spatial Anchor} node to generate a 3D spatial anchor. For more detailed step-by-step scenario, please refer to the supplementary video.}


%% file: contents/Evaluation.tex
\section{Evaluation}


We conducted a user study with twelve participants to evaluate \xrops' usability. 
The study included a training session and five demo tasks, with the time for each step recorded.  
Usability, strengths, and weaknesses were assessed using the System Usability Scale (SUS~\cite{bangor2008empirical}) survey and a post-interview.



%
%
%
\subsection{User Study Design}

\noindent\textbf{Participants:} We recruited twelve participants (P1-12) from varied backgrounds. P1, 2, 4, and 11 are general users with limited programming experience from fields like marketing, physics, biomedical engineering, and neuroscience research. P3, 5, 6, and 10 have programming experience in visualization, but are not experts in XR. P7, 8, 9, and 12 are professional XR programmers with direct involvement in XR visualization research.
%
%
%
%
%
\vspace{5pt}

\noindent\textbf{Study procedure:} Participants first completed a pre-questionnaire covering demographics, XR experience, and visualization programming knowledge.
Then, they are trained on HoloLens2 and \xrops~visualizing the Iris flower dataset~\cite{fisher1936use}. 
After training, participants completed the five demo tasks sequentially as outlined in Table \ref{tasks}. 
Afterward, they filled out the SUS questionnaire on \xrops~and provided feedback on its strengths, weakness, potential applications, and areas for improvement, alongside their previous XR and data visualization experiences. 
%
%
%
%
%
%
%
%


\begin{table}[]
\fontsize{6pt}{6pt}\selectfont
\caption{A brief introduction to the five demo tasks used in the user study.}
\label{tasks}
\setlength{\tabcolsep}{0.5em} %
\renewcommand\baselinestretch{1.3}
\renewcommand{\arraystretch}{2}
\begin{tabular}{|M{0.05\columnwidth}|m{0.9\columnwidth}|}
\hline
\sffamily{Task} & \sffamily{Description} \\
\hline
\hline
                \sffamily{T1} & \sffamily{The participants created the IA environment of \textbf{S2}.}  \\ \hline
                 \sffamily{T2} & \sffamily{The participants performed visual analytics of the car dataset and found of three pieces of information, such as "the origin of the cars with the highest miles per gallon is japan". The attributes of the dataset consist of car name, miles per gallon, number of cylinders, displacement, horsepower, weight in lbs, acceleration, year of release, and origin.} 
  \\ \hline
                 \sffamily{T3} & \sffamily{The participants visualized of Stanford Bunny [1] mesh dataset.}  \\ \hline
                  \sffamily{T4} & \sffamily{The participants compared three different cryptocurrencies, such as Bitcoin, Ripple, and Ethereum (\textbf{S1}).}    \\ \hline
                 \sffamily{T5} & \sffamily{The participants performed \textbf{S3} using a dummy head}    \\ \hline

\end{tabular}
\end{table}



\subsection{User Study Results}


%
%
Participants, regardless of their XR or programming experience, successfully created the desired IA environment using \xrops.
We evaluated usability with the SUS questionnaire, a reliable tool across various systems~\cite{bangor2008empirical} even with few participants. 
It consists of ten questions rated on a five-point Likert scale (1=strongly disagree, 5=strongly agree). 
%
%
%
%
Results show \xrops~has \textit{good} usability, with an average SUS score of 71.67 \minor{(See Supplementary Fig.~S3)}.
We also recorded each duration of the training session, T1, and T2 for all participants (note that T3-5 were not goal-oriented, so time was not recorded). 
The average duration for the training session was 17 minutes and 54 seconds ($\pm$ 6 minutes 33 seconds).
The average duration for T1 was 16 minutes 35 seconds ($\pm$ 9 minutes 11 seconds), and for T2 was 10 minutes 16 seconds ($\pm$ 6 minutes 23 seconds). 
%
Notably, participants with XR development experience tended to explore more options and spend more time on tasks.

\subsubsection{Feedback}


\noindent\textbf{Strength:} 
Most participants found \xrops~user-friendly and accessible, even for beginners, thanks to its visual programming interface. 
They reported that creating and modifying workflows by connecting nodes were intuitive and easy to manage. 
Participants P3, P10, and P11 noted that the Vis Linking node made comparing multiple visualizations simple. 
P2 and P10 appreciated the ability to customize plots with the text editor. 
%
%
P1 found real-time data analysis helpful, while P8 noted that \xrops~enabled quick data checks.

\vspace{5pt}
\noindent\textbf{Weakness:} 
P5 and P8 stated that they were unfamiliar with the encoding method that maps data attributes to plot axes. 
\edit{In the case of P2 and P5, the drawback of visual programming is that it is not possible to freely modify the node, which enables easy authoring but seems to have limitations in flexible modification. P2 suggested that the shortcomings should be solved by diversifying nodes to create more specific applications.}
\edit{P8 reported fatigue from having to switch screens to perform even simple tasks through the 2D \ui.}
\vspace{5pt}
\noindent\textbf{Room for improvement:} 
P8 and P12 suggested it would be more convenient if GUI interaction and visualization coding could occur directly within the XR space. 
P7 recommended adding error displays during a workflow creation for easier debugging.
P8 also suggested consistency between 2D previews and XR visualizations.

%% file: contents/Discussion.tex
\subsection{Discussion}
\label{sec:discussion}

\textbf{Dynamic Workflow Management:} 
 \new{
\xrops~manages workflow components such as data processing, development, visual encoding, and deployment within a single system. 
While previous IA toolkits have introduced valuable concepts such as rapid feedback, reactive programming, and hot-reloading, the scope of reactivity in these approaches has been limited, mainly focusing on changes to data sources or encoding modifications. 
%
In contrast, \xrops~handles adjustments to algorithms, data generation through interaction, and the integration of collaborators.
Through the demonstration of use cases, \xrops~has proven that it extends beyond the scope of traditional IA toolkits by enabling dynamic workflow modifications in response to user activities, thereby fostering a dynamic and responsive development environment.
}

\vspace{5pt}
\noindent\textbf{Visual Programming:} 
\new{\xrops~employs visual programming for workflow management, known for its intuitive nature and adaptability across educational settings. 
This approach suits diverse tasks better than traditional methods like JSON scripting, catering to users with varied backgrounds, even those without web development or visualization experience. 
However, visual programming may lack the flexibility of traditional programming and can pose challenges with complex node connections. 
To address these limitations, \xrops~provides task abstractions based on common usage 
and allows custom nodes for user code. 
Feedback during the user study from users with diverse backgrounds confirms the framework's ease of use, with even non-programmers quickly mastering it during training and task execution.}


\vspace{5pt}
\noindent\textbf{Data Versatility and Sensor usage:} 
\new{One of the key advantages of IA lies in its capability to capture the user's surrounding environment, enabling data analysis based on spatial awareness. This functionality is directly linked to various application areas, including user interaction, mapping between the environment and data, large-scale data analysis, and the three-dimensional comprehension of scientific data. }
However, we observed that existing authoring toolkits are insufficiently accommodating these requirements. 
The majority of existing toolkits are focused on tabular data visualization and are notably lacking in preparations for the visualization of scientific or dynamically changing data. 
\new{Although there have been attempts to utilize environmental data~\cite{fleck2022ragrug, chen2023iarvis}, there remains a need for authoring capabilities that facilitate the creation of a much wider range of scenarios.
\xrops~addresses this gap by enabling the exploration of new data types and methods, thereby fully leveraging the potential of IA.}

\subsection{Limitations}

\new{\xrops~introduces web-based visual programming for IA development, which can reduce the user’s programming effort during the development process. 
However, as the number of nodes in the workflow grows, the dataflow graph can become cluttered and dense. 
This can be avoided by grouping common tasks into one custom node. 
Another possible solution is a hierarchical user interface (e.g., multi-scale graph visualization), though it was outside the scope of this work.
\xrops~shows potential in implementing scientific data analysis via an IA authoring toolkit, handling diverse data types like meshes, volumes, point clouds, and images, unlike many other IA authoring tools. However, being in an early stage, further development of scientific visualization features is needed to support a broader range of visualizations. This includes improving rendering techniques for volume, mesh, and point cloud data, as well as enabling real-time interaction with volumes. Future research is essential to address performance challenges, particularly in rendering large-scale volume data on standalone XR devices, given the current lack of specific optimization technology for data scalability. 
\xrops's server-client architecture holds potential for improving performance in massive data visualization through remote rendering.}

%% file: contents/Conclusion.tex
\section{Conclusion and future work}

In this paper, we introduced \xrops, a web-based IA authoring system designed to enable flexible and dynamic workflow modifications on the fly, ensuring a seamless development experience in IA.
%
%
%
%
%
With \xrops' intuitive visual programming interface, even those without programming experience can swiftly construct their desired XR environments with minimal training. 
\xrops~is also equipped with essential functionalities to facilitate the integration of sensor data from XR devices, enabling users to interact with their environments with ease. 
We believe that \xrops~marks a significant step forward in creating a more accessible and user-friendly XR authoring system, which has the potential to expand the scope of XR technologies to a larger audience.


Although the current system provides functional nodes for various immersive analytics scenarios, we plan to continuously develop and expand the collection of functional nodes that can be applied to various real-world scenarios not covered in this paper. 
%
\new{We also plan to extend the current system to integrate recent XR environments, such as WebXR~\cite{webxr}.}
%

%
%

